\documentclass[aps,pra,floatfix,groupedaddress,footinbib,notitlepage,showpacs,superscriptaddress]{revtex4-1}
\usepackage{bm,bbm,bbold,graphicx,graphics,amssymb,amsmath,amsfonts,dsfont,color}

\newcommand{\ket}[1]{|#1\rangle}

\newcommand{\ketbra}[2]{|#1\rangle\langle #2|}
\newcommand{\mrm}[1]{\mathrm{#1}}

\newcommand{\average}[1]{\left\langle#1\right\rangle}

\newcommand{\ignore}[1]{}

\let\oldsqrt\sqrt
\def\sqrt{\mathpalette\DHLhksqrt}
\def\DHLhksqrt#1#2{%
\setbox0=\hbox{$#1\oldsqrt{#2\,}$}\dimen0=\ht0
\advance\dimen0-0.2\ht0
\setbox2=\hbox{\vrule height\ht0 depth -\dimen0}%
{\box0\lower0.4pt\box2}}

\DeclareFontFamily{OT1}{pzc}{}
\DeclareFontShape{OT1}{pzc}{m}{it}%
              {<-> s * [1.25] pzcmi7t}{}
\DeclareMathAlphabet{\mathpzc}{OT1}{pzc}%
                                 {m}{it}

\begin{document}

\title{Entropy production and non-Markovian dynamical maps}

\author{S. Marcantoni}
\email{stefano.marcantoni@ts.infn.it}
\affiliation{Department of Physics, University of Trieste, I-34151 Trieste, Italy}
\affiliation{National Institute for Nuclear Physics (INFN), Trieste Section, I-34151 Trieste, Italy}

\author{S. Alipour}
\affiliation{School of Nano Science, Institute for Research in Fundamental Sciences (IPM), Tehran 19538, Iran}

\author{F. Benatti}
\email{benatti@ts.infn.it}
\affiliation{Department of Physics, University of Trieste, I-34151 Trieste, Italy}
\affiliation{National Institute for Nuclear Physics (INFN), Trieste Section, I-34151 Trieste, Italy}

\author{R. Floreanini}
\affiliation{National Institute for Nuclear Physics (INFN), Trieste Section, I-34151 Trieste, Italy}

\author{A. T. Rezakhani}
\affiliation{Department of Physics, Sharif University of Technology, Tehran 14588, Iran}

\begin{abstract}
In the weak-coupling limit approach to open quantum systems, the presence of the bath is eliminated and accounted for by a master equation that introduces dissipative contributions to the system reduced dynamics. 
Within this framework, there are no bath entropy contributions to the entropy balance. We show that, as a consequence, the entropy production fails to be positive for a class of physically legitimate 
(i.e., completely positive and trace preserving) non-Markovian dynamical maps. Moreover, in the absence of the semigroup property, if the reduced dynamics has a thermal asymptotic state, this need not be stationary. 
In this case, even the integrated entropy production becomes negative. These observations imply that, when the conditions leading to reduced dynamics of semigroup type are relaxed, a consistent formulation of the second law of thermodynamics requires that the environment contribution to the entropy balance be explicitly taken into account. 
\end{abstract}

\flushbottom
\maketitle

\thispagestyle{empty}

\section*{Introduction}

Since the late seventies the laws of thermodynamics have been formulated for an open quantum system interacting with a reservoir (called also alternatively, ``environment" or ``bath") in equilibrium at inverse temperature $\beta$ using the theory of quantum dynamical semigroups \cite{Alicki_book}. In this approach, a reduced dynamics for an open quantum system is obtained through 
the so-called weak-coupling limit techniques based on three assumptions: 
1) that the environment be weakly coupled to the system of interest, 2) that the initial common state be factorized, and 3) that the time scales of the system and environment be clearly separated so that a Markovian approximation is feasible. It follows that the reduced dynamics consists of a semigroup of trace-preserving completely-positive dynamical maps affecting only the degrees of freedom of the system; 
the presence of the bath is eliminated and accounted for by a dissipative modification of the master equation that effectively embodies bath-induced noisy and damping effects. Within this framework, it has been shown that the entropy production, defined in analogy with classical irreversible thermodynamics \cite{Book:deGroot} as the difference between the total entropy variation and the entropy flux due to the heat exchange with the environment, can be related to the time-derivative of the relative entropy and is always nonnegative \cite{Alicki-79,Spohn}. This important property has been proposed as a statement of the second law of thermodynamics in a context where no entropic bath terms contribute directly to the entropy balance. However, this formulation of the second law relies on both the semigroup composition law and on the assumption of a thermal asymptotic state. Although a more general formulation can be given for a quantum dynamical semigroup relaxing to a state that is not in the Gibbs form \cite{entroSpohn}, in this case a direct thermodynamic interpretation of the relative entropy is not available.

In the last decade, non-Markovian dynamics of open quantum systems received considerable attention since a realistic description of many physical open quantum systems in interaction with their environment requires to relax assumption 3 above, thus renouncing the usual Markovian approximation \cite{review,RHPreview,deVega}. Even though the very definition of Markovianity in the quantum domain is still debated, in the following we adopt the point of view of Refs. \cite{RHP,Chruscinski} associating a non-Markovian behavior with the lack of a property known as ``CP-divisibility."

The thermodynamics of an open quantum system experiencing a non-Markovian time evolution is an interesting subject of current research \cite{Fazio,Bylicka}. In particular, it is worth studying whether the second law of thermodynamics can be derived from the properties of the dynamical maps as in the Markovian case. It has been argued in Refs. \cite{Argentieri1,Argentieri2} that non-completely-positive (non-CP) dynamics can lead to a negative entropy production and hence to a violation of the second law of thermodynamics. We, however, show in this paper that a negative entropy production can also occur for a class of CP (thus physically legitimate) non-Markovian dynamics. We argue that this interesting outcome should not be interpreted as a violation of the second law of thermodynamics but as an evidence that, by relaxing the conditions leading to reduced dynamical maps of semigroup type, a proper formulation of the laws of thermodynamics can only be obtained by explicitly dealing with the environment. 
As a consequence, the second law of thermodynamics must be expressed in terms of the sum of the variations of the entropy of both open quantum system and environment which, in the absence of initial correlations between the two systems is known to be non-negative \cite{book:Peres,Esposito,Sagawa}. Whether this argument also extends to more general approaches that go beyond the weak-coupling limit, for instance by considering  orrelations in the bath higher than the second order \cite{Modi}, is an interesting open problem, deserving a separate investigation.

In the following, we first review the standard thermodynamic description of open quantum systems in the Markovian case, with particular emphasis on what has been proposed as a statement of the second law. Then, we briefly present the basic features of non-Markovian dynamical maps. We concentrate on evolutions that thermalize the system to a unique Gibbs state, so that we can write the entropy production by means of a well-defined reference temperature.  Subsequently, we firstly discuss an example of non-Markovian dynamics in which the entropy production is not always positive and an example showing that the integrated entropy production can be negative too if the asymptotic thermal state is not an invariant state for the dynamics. This can happen when the semigroup property does not hold \cite{asymptotic}. Finally, we comment on when the necessary positivity of the entropy balance of open system and bath together leads to the positivity of the entropy production for the open system alone, showing that the connection can fail by means of a third example.
The paper is concluded with some final remarks.

\section*{Thermodynamics of an open quantum system}

\label{sec:ther}

Consider a (possibly driven) open quantum system with (a possibly time-dependent) Hamiltonian $H_{\tau}$ described by a finite-dimensional Hilbert space $\mathpzc{H}$, whose state at time $\tau$ (where $\tau\geqslant0$) is given by $\varrho_{\tau}$. 
The internal energy is given by
\begin{equation}
\mathds{U}_{\tau} :=\mrm{Tr}\left[ \varrho_{\tau} H_{\tau}\right],
\end{equation}
and one can distinguish the \textit{heat} and \textit{work} contributions to its time variation ($\partial_{\tau} \mathds{U}_{\tau}$) as follows \cite{Alicki-79}:
\begin{align}
\partial_{\tau} \mathds{W}_{\tau} &:=\mrm{Tr}\left[ \varrho_{\tau}\,\partial_{\tau} H_{\tau}\right], \label{work}
\\
\partial_{\tau} \mathds{Q}_{\tau} &:=\mrm{Tr}\left[ \partial_{\tau}\varrho_{\tau}\, H_{\tau}\right]. \label{heat}
\end{align} 
This is a reasonable choice since the work power vanishes if the Hamiltonian is time-independent, namely if there is no external driving; whereas the heat flux is zero when the system is isolated from any kind of environment and thus evolves according to the Schr\"odinger time evolution generated by $H_\tau$. In the following, we concentrate on undriven open quantum systems (where $H_\tau=H$) exchanging heat with their environment, which is taken to be a heat bath at inverse temperature $\beta$ (note that an explicitly time-dependent Hamiltonian can always be considered by extending the formalism as done in Ref. \cite{Alicki-79}).

Concerning the entropy balance, one can use the von Neumann entropy $\mathds{S}$ to describe the total entropy of the system out of equilibrium and define the entropy production  $\bm{\sigma}_{\tau}$ in analogy with classical irreversible thermodynamics, 
\begin{align}
\mathds{S}_{\tau} &:= -\mrm{Tr}\left[ \varrho_{\tau} \log\varrho_{\tau}\right], \\
\bm{\sigma}_{\tau} &:= \partial_{\tau} \mathds{S}_{\tau} -\beta \partial_{\tau} \mathds{Q}_{\tau}.
\end{align}
Throughout the paper we assume $k_{\mathrm{B}}\equiv\hbar\equiv 1$. A straightforward calculation shows that $\bm{\sigma}_{\tau}$ can be conveniently rewritten in terms of the derivative of the relative entropy between the state $\varrho_{\tau}$ and the Gibbs state $\varrho^{(\beta)}= \mathrm{e}^{-\beta H}/\mathrm{Tr}[\mathrm{e}^{-\beta H}]$, \textit{i.e.}, 
\begin{equation}
\label{sigrel}
\bm{\sigma}_{\tau} = -\partial_{\tau} \mathds{S}(\varrho_{\tau}|\hskip-0.5mm|\varrho^{(\beta)}), 
\end{equation}
where $\mathds{S}(\varrho|\hskip-0.5mm|\varrho^{\prime}):= \mrm{Tr}\left[\varrho \log\varrho - \varrho \log \varrho^{\prime}\right]$.
Equation \eqref{sigrel} holds provided that the Hamiltonian is time-independent and that the environment is a heat bath in thermal equilibrium, without other dynamical assumptions. 

If the reduced dynamics of the open quantum system is described by a master equation in the Lindblad form such that the unique asymptotic state is a Gibbs thermal state at the heat bath temperature,
\begin{equation}
\partial_{\tau} \varrho_{\tau}=-i[H,\varrho_{\tau}] + \mathpzc{L}[\varrho_{\tau}] ,\qquad
\mathpzc{L}[\varrho_{\tau}]= \sum_k \Big( V_k\varrho_{\tau}V_k^{\dagger} - \frac{1}{2}\big\lbrace V_k^{\dagger}V_k,\varrho_{\tau} \big\rbrace \Big),\qquad
\label{Lind}
\lim_{\tau \rightarrow \infty}\varrho_{\tau}=\varrho^{(\beta)},
\end{equation}
one can consistently express the second law of thermodynamics through the nonnegativity of the entropy production $\bm{\sigma}_{\tau} \geqslant 0$ \cite{Alicki-79}. The proof is based on the fact that any asymptotic state is necessarily also an invariant state for the dynamics due to the semigroup property (given $\varrho_{\tau}=\Lambda_{\tau}[\varrho_0]$, one has $\Lambda_{\tau+\delta}=\Lambda_{\tau}\Lambda_{\delta}$) and that the relative entropy is decreasing under CP maps \cite{book:Nielsen},
\begin{equation}
\label{ineq}
\partial_{\tau} \mathds{S}\big(\Lambda_{\tau} [\varrho_0] |\hskip-0.5mm| \varrho^{(\beta)}\big)=\partial_{\tau} \mathds{S}\big(\Lambda_{\tau} [\varrho_0] |\hskip-0.5mm| \Lambda_{\tau} [\varrho^{(\beta)}]\big) = \lim_{\delta \to 0^+}\frac{\mathds{S}\big(\Lambda_{\delta}\Lambda_{\tau} [\varrho_0] |\hskip-0.5mm| \Lambda_{\delta}\Lambda_{\tau} [\varrho^{(\beta)}]\big) -\mathds{S}\big(\Lambda_{\tau} [\varrho_0] |\hskip-0.5mm| \Lambda_{\tau} [\varrho^{(\beta)}]\big) }{\delta} \leqslant 0.
\end{equation}

In the above approach the dynamics of the open quantum system is dissipative due to the presence of a suitable environment. However, its presence is not explicitly taken into account in the two definitions \eqref{work} and \eqref{heat}. A different perspective was recently considered towards a formulation of thermodynamics of two interacting quantum systems none of which can be neglected \cite{Alipour}. In this case, the heat balance relation strongly depends on the correlations between the two parties built up through the interactions. We shall use this approach in Example III.

\section*{Non-Markovian dynamical maps}

\label{sec: nonMarkov}

Recently, the study of non-Markovian quantum dynamical maps has received much attention because of the high degree of control reached in many experimental setups that allows to exploit physical effects not explainable with the use of a quantum dynamical semigroup. Although various approaches exist in the literature, a general formulation of non-Markovianity is still under debate 
\cite{review}. 

In this work, we use the definition adopted in Ref. \cite{RHP}, where the non-Markovianity is associated with the lack of CP-divisibility of a dynamical map. A (CP and trace-preserving) dynamical map $\Lambda_{\tau}$ is called CP-divisible if one can write 
\begin{equation}
\Lambda_{\tau}= \mathpzc{V}_{\tau,s}\Lambda_s, \qquad 0\leqslant s \leqslant\tau,
\end{equation}
such that the intertwining map $\mathpzc{V}_{\tau,s}$ is CP for all $\tau, s$. The quantum dynamical semigroup generated by the Lindblad master equation \eqref{Lind} obviously satisfies this property because there we have $\mathpzc{V}_{\tau,s}=\Lambda_{\tau-s}$. 
When $\mathpzc{V}_{\tau,s}$ is positive, the map $\Lambda_{\tau}$ is called P-divisible (which is weaker than CP-divisibility). Most non-Markovianity measures are based on P-divisibility \cite{review}. 

Following Ref. \cite{degree}, we call a dynamical map which is not even P-divisible an \textit{essentially} non-Markovian map.

In order to have a meaningful thermodynamic interpretation of this kind of dynamics and to compare it with the situation described in the previous section, we restrict to those evolutions that have a Gibbs state $\varrho^{(\beta)}$ as their unique asymptotic state. In this case, one can use $\beta^{-1}$ as a reference equilibrium temperature and the entropy production $\bm{\sigma}_{\tau}$ reads as in Eq. \eqref{sigrel}. For a non-Markovian evolution the asymptotic state is not necessarily an invariant state of the dynamics \cite{asymptotic}, thus we can distinguish two different situations,
\begin{itemize}
\item[(i)] 
$\forall \tau\quad \Lambda_{\tau} [\varrho^{(\beta)}] = \varrho^{(\beta)}$,
\item[(ii)] $\exists \tau
\quad \text{such that}\quad\Lambda_{\tau} [\varrho^{(\beta)}] \neq \varrho^{(\beta)}$.
\end{itemize}
In the first case, since $\Lambda_\tau$ is always taken to be CP, the integrated entropy production $\bm{\Sigma}_{\tau}:= \int_0^{\tau}\! \bm{\sigma}_{\tau'} \,\mrm{d}\tau' $ 
is always nonnegative. Indeed, by means of Eq.~\eqref{sigrel}, one obtains 
\begin{equation}
\bm{\Sigma}_{\tau}=  \mathds{S}(\varrho_0 |\hskip-0.5mm| \varrho^{(\beta)}) - \mathds{S}(\Lambda_{\tau} [\varrho_0] |\hskip-0.5mm| \varrho^{(\beta)}) 
=  \mathds{S}(\varrho_0 |\hskip-0.5mm| \varrho^{(\beta)})- \mathds{S}(\Lambda_{\tau} [\varrho_0] |\hskip-0.5mm|\Lambda_{\tau} [\varrho^{(\beta)}]) \geqslant 0,
\label{Sigma}
\end{equation}
where we have used that the relative entropy monotonically decreases under completely positive maps and property (i). 
Note, however, that the rate $\bm{\sigma}_{\tau}$ can become temporarily negative if the dynamics is essentially non-Markovian (\textit{i.e.}, not P-divisible). Instead, 
it has been recently proved that if $\Lambda_{\tau}$ is at least P-divisible, 
then 
$\partial_{\tau} \mathds{S}\left(\Lambda_{\tau} [\varrho_1] |\hskip-0.5mm| \Lambda_{\tau} [\varrho_2] \right) \leqslant 0$, 
for any pair of density matrices $\varrho_1$ and $\varrho_2$ \cite{positive};  in which case $\bm{\sigma}_{\tau}\geqslant 0~\forall \tau$.  
Concerning the lack of P-divisibility, in Example I, we discuss a dynamics  which fulfills property (i) but with  
$\bm{\sigma}_{\tau}<0$ in a certain time interval.

In case (ii) the above line of argumentation cannot be used to show that $\bm{\Sigma}_{\tau} \geqslant0$
because the necessary substitution $\varrho^{(\beta)}\to\Lambda_{\tau} [\varrho^{(\beta)}]$
is not allowed. 
In fact, in Example II, we show that the inequality in Eq. \eqref{Sigma} may be violated.

We argue that in a non-Markovian context a possible negative entropy production 
is not directly associated with a violation of the second law of thermodynamics. Rather, it indicates that the presence of the environment at the origin of the dissipative dynamics cannot be entirely neglected. This point of view is also supported by the characterization of non-Markovianity in terms of a backflow of information from the environment to the system. Indeed, lack of P-divisibility can make the distinguishability of two states of the system increase in time \cite{review}. One may then 
relate such a behavior to processes that cause the entropy of the environment to increase. In fact, the main purpose of this work is to motivate and support the point of view that a proper formulation of the second law of thermodynamics for a non-Markovian open quantum system cannot be based only on its reduced dynamical maps. In this respect, it seems better to follow the approach of Refs. \cite{Esposito,Alipour} and consider explicitly the reservoir in the entropy balance---as we will explicitly do in Example III.

\section*{Example I: Qubit in a thermal bath}
\label{sec: example1}

As a first example, we consider the following master equation \cite{Fazio}:
\begin{equation}
\partial_{\tau}\varrho_{\tau}=-i\left[\frac{\omega}{2}\sigma_z,\varrho_{\tau} \right] + \frac{\gamma_{\tau}(n+1)}{2} 
\big(2\sigma_{-}\varrho_{\tau} \sigma_{+} -\{\sigma_{+}\sigma_{-},\varrho_{\tau}\} \big) 
+ \frac{\gamma_{\tau}n}{2} \big(2\sigma_{+}\varrho_{\tau} \sigma_{-} -\{\sigma_{-}\sigma_{+},\varrho_{\tau}\} \big),
\label{master1}
\end{equation}
where $n= (\mrm{e}^{\beta \omega} -1)^{-1}$, $\gamma_{\tau}$ is a time-dependent damping rate, and $\sigma_{a}$ ($a\in\{x,y,z\}$) are the Pauli matrices (with $\sigma_{\pm}=\sigma_{x}\pm i\sigma_{y}$). By choosing a constant damping, we can readily recover the usual Lindblad master equation for a qubit interacting with a thermal bath at inverse temperature $\beta$. One can show that Eq.~\eqref{master1} generates a CP dynamical map $\Lambda_{\tau}$ iff $\int_0^{\tau}\gamma_{\tau'}\,\mrm{d}\tau' \geqslant 0$ (see Theorem 3.1 in Ref. \cite{commutative}). Moreover, $\Lambda_{\tau}$ is both CP-divisible and P-divisible iff $\gamma_{\tau}\geqslant 0$ \cite{degree}. 

By means of the Bloch representation 
\begin{equation}
\label{bloch}
\varrho=\frac{1}{2}\left(1+x\sigma_x +y\sigma_y+z\sigma_z\right)
\end{equation}
of $\varrho$, the solution of Eq.~\eqref{master1} in terms of the Bloch vector components, $(x,y,z)$, reads
\begin{align}
&x_{\tau}\pm iy_{\tau}= \mrm{e}^{-\Gamma_{\tau}\pm i\omega \tau}( x_{0}\pm iy_{0}), \\
&z_{\tau}= \mrm{e}^{-2\Gamma_{\tau}} (z_{0}-z_{\infty})+ z_{\infty},
\end{align}
where $\Gamma_{\tau}=(1/2)\coth(\beta\omega/2) \int_0^{\tau}\gamma_{\tau'}\,\mrm{d}\tau'$ and $z_{\infty}=-\tanh(\beta\omega/2)$. Note that the Gibbs state is an invariant state of the dynamics and it is also the unique asymptotic state provided that $\lim_{\tau \to \infty}\Gamma_{\tau}=\infty$. Hence, the integrated entropy production $\bm{\Sigma}_{\tau}$ is nonnegative because of Eq. \eqref{Sigma}. Nevertheless, we could expect the entropy production to become transiently negative when the dynamics fails to be P-divisible, {\it i.e.,} becoming essentially non-Markovian. This is indeed the case as we show in the following.

A straightforward calculation yields the heat flux as
\begin{equation}
\partial_{\tau} \mathds{Q}_{\tau} = \frac{\omega}{2}\partial_{\tau} z_{\tau}= -\frac{\omega}{2}\gamma_{\tau}\coth(\beta\omega/2)\, \mrm{e}^{-2\Gamma_{\tau}}[z_{0}+\tanh(\beta\omega/2)],
\end{equation}
so that its sign depends both on the initial condition and on the instantaneous rate $\gamma_{\tau}$. The entropy variation is written by means of the eigenvalues $(1\pm r_{\tau})/2$ of the density matrix as
\begin{equation}
\label{entro}
\partial_{\tau}\mathds{S}_{\tau} = -\frac{1}{2}\log\!\left(\frac{1+r_{\tau}}{1-r_{\tau}}\right)\,\partial_{\tau} r_{\tau}=\frac{\gamma_{\tau}\coth(\beta\omega/2)}{4r_{\tau}}~\log\!\left(\frac{1+r_{\tau}}{1-r_{\tau}}\right)\left[x^2_{\tau}+y^2_{\tau}+2z^2_{\tau}+2z_{\tau}\tanh(\beta\omega/2)\right],
\end{equation}
where $r_{\tau}=\sqrt{x^2_{\tau}+y^2_{\tau}+z^2_{\tau}}$, and the sign of $\partial_{\tau}\mathds{S}_{\tau}$ 
again depends on the rate $\gamma_{\tau}$ and on the initial condition, as one can see rewriting the term in the last bracket as
\begin{equation}
x^2_{\tau}+y^2_{\tau}+2z^2_{\tau}+2z_{\tau}\tanh(\beta\omega/2)=\mrm{e}^{-4\Gamma_{\tau}}(z_{0}-z_{\infty})^2 + \mrm{e}^{-2\Gamma_{\tau}}\left[ x^2_{0}+y^2_{0} + (z_{0}-z_{\infty})z_{\infty} \right].
\end{equation}
The entropy production  thus reads
\begin{equation}
\label{entroprod}
\hskip-.3cm
\bm{\sigma}_{\tau}=\gamma_{\tau}\coth(\beta\omega/2)\mrm{e}^{-2\Gamma_{\tau}}\left[ \left(x^2_{0}+y^2_{0}+2\mrm{e}^{-2\Gamma_{\tau}}[z_{0}+|z_{\infty}|]^2  \right)\frac{1}{4r_{\tau}}\log\left(\frac{1+r_{\tau}}{1-r_{\tau}}\right)+ (z_{0}+|z_{\infty}|) \left( \frac{\beta\omega}{2}- \frac{|z_{\infty}|}{2r_{\tau}}\log\left(\frac{1+r_{\tau}}{1-r_{\tau}}\right) \right) \right].
\end{equation}
We prove in the Supplementary Information that the expression in the square brackets above is always positive, whence the sign of $\bm{\sigma}_{\tau}$ corresponds to the sign of $\gamma_{\tau}$. 
In other words, 
whenever the damping rate is negative (so that the dynamics is essentially non-Markovian), the entropy production becomes negative too.

We stress the fact that a physically legitimate dynamics, namely CP and trace-preserving, can lead to a negative entropy production. This property is associated with the lack of P-divisibility, that is, it arises in the class of essential non-Markovian maps.

\section*{Example II: Qubit amplitude damping channel}

\label{sec: example2}

This second example aims at highlighting the role of the existence of an asymptotic non-invariant state concerning the entropy production. Consider a generalized amplitude damping channel $\Phi[\,\cdot\,]= \sum_{i=0}^{3} E_i (\cdot) E_i^{\dagger}$ where 
\begin{eqnarray}
E_0&=&\sqrt{p}\left( \ketbra{0}{0} + \sqrt{1-\gamma} \ketbra{1}{1} \right)\ ,\quad E_1= \sqrt{p\gamma}\ketbra{0}{1}\ ,\nonumber\\ 
E_2&=&\sqrt{1-p}\left( \sqrt{1-\gamma}\ketbra{0}{0} + \ketbra{1}{1} \right)\ ,\quad
E_3= \sqrt{(1-p)\gamma}\ketbra{1}{0}\ , 
\end{eqnarray}
and $p,\gamma \in [0,1]$ \cite{book:Nielsen}. Adjusting the parameters $p_{\tau}$ and $\gamma_{\tau}$ as suitable functions of time, 
one can construct a physically legitimate dynamics namely a one-parameter family of CP and trace-preserving maps $\Phi_{\tau}$ as
\begin{equation}
\varrho_{0}=\frac{1}{2}\left(1+x\sigma_x +y\sigma_y+z\sigma_z\right) \mapsto 
\Phi_{\tau}[\varrho_{0}]=
\frac{1}{2}\left(1+x_{\tau}\sigma_x +y_{\tau}\sigma_y+z_{\tau}\sigma_z\right)\ ,
\end{equation}
where the Bloch representation~\eqref{bloch} of the density matrix has been used. Explicitly, the Bloch vector components at time $\tau$ read
\begin{equation}
x_{\tau}\pm iy_{\tau}= \sqrt{1-\gamma_{\tau}}(x_{0}\pm iy_{0})\ ,\quad 
z_{\tau} = -\gamma_{\tau} + 2p_{\tau}\gamma_{\tau} + z_{0}(1-\gamma_{\tau})\ .
\end{equation} 
We can impose a unique asymptotic state to exist for this family of dynamical maps by means of the condition $\gamma_{\infty}=1$; moreover, it is a Gibbs state $\varrho^{(\beta)}= \mrm{e}^{-\beta \sigma_z}/\mrm{Tr}\big[\mrm{e}^{-\beta \sigma_z}\big]$, 
if the further condition $2p_{\infty}-1=-\tanh(\beta)$ is fulfilled. The initial condition instead implies that $\gamma_{0}=0$. We can choose the time dependence of $p$ and $\gamma$ such that they become compatible with all these constraints. A possibility is to set
\begin{equation}
2p_{\tau}-1= \mrm{e}^{-\epsilon \tau}\sin^2(\epsilon \tau)-\tanh(\beta)\ ,\quad
\gamma_{\tau} =1-\mrm{e}^{-2\lambda \tau}\ ,
\end{equation}
so that a quantum dynamical semigroup is recovered for $\epsilon=0$. This can be seen from the time-dependent generator of $\Phi_{\tau}$,
\begin{equation}
\mathpzc{L}_{\tau}[\,\cdot\,]=a^{(-)}_{\tau} \Big( \sigma_-(\cdot)\sigma_+ -\frac{1}{2}\left\lbrace \sigma_+ \sigma_-, \cdot \right\rbrace \Big)\,+\, a^{(+)}_{\tau}\Big( \sigma_+(\cdot)\sigma_- -\frac{1}{2}\left\lbrace \sigma_- \sigma_+, \cdot \right\rbrace \Big)\ ,
\end{equation}
where 
\begin{equation}
a^{(\pm)}_{\tau}= \frac{1}{4}\left(\frac{p^{(\pm)}_{\tau}}{1-\gamma_{\tau}}\partial_{\tau}\gamma_{\tau}\pm \gamma_{\tau} \partial_{\tau} p_{\tau}\right)\ ,
\end{equation}
with $p^{(-)}_{\tau}=1-p_{\tau}$ and $p^{(+)}_{\tau}=p_{\tau}$, which becomes a time-independent Lindbladian in the limit $\epsilon\to 0$. 

\begin{figure}
\flushright
\includegraphics[scale=.4]{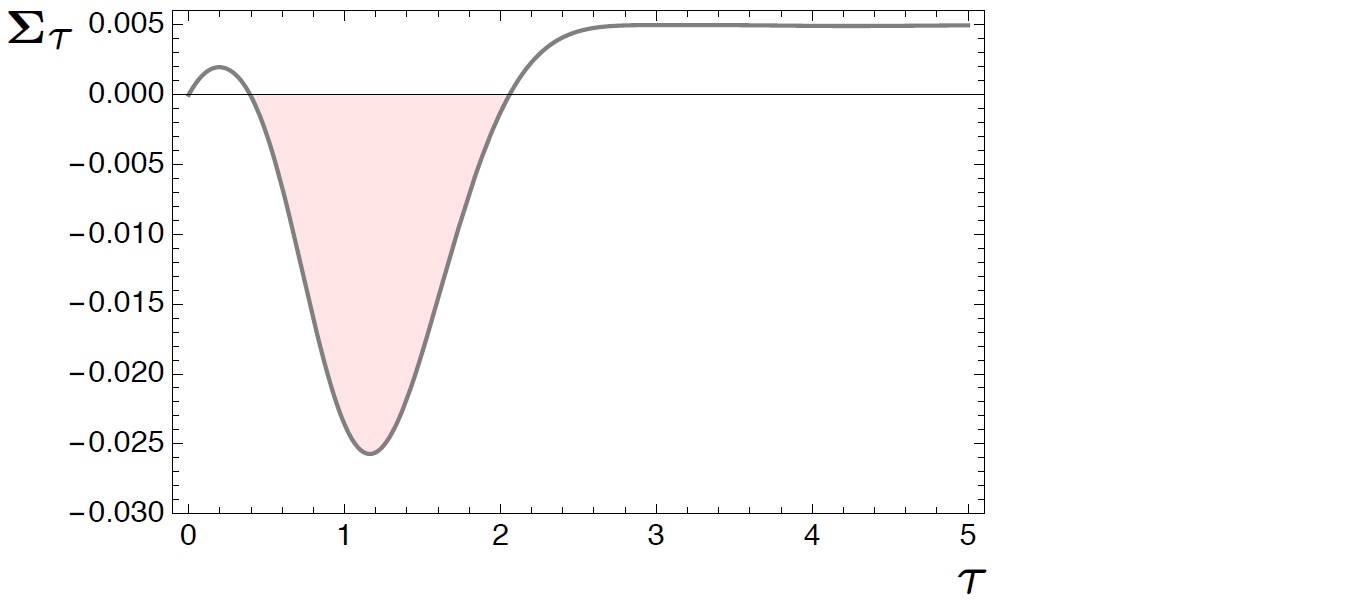}
\caption{Transient negativity of the integrated entropy production  
$\bm{\Sigma}_{\tau}$.}
\label{fig:1}
\end{figure}

The quantity of interest is the difference between the relative entropies [Eq. \eqref{Sigma}], 
\begin{equation}
\label{diff}
\bm{\Sigma}_{\tau} =\mathds{S}(\varrho_0|\hskip-0.5mm|\varrho^{(\beta)})- \mathds{S}(\varrho_{\tau}|\hskip-0.5mm|\varrho^{(\beta)})=-\frac{1}{2}\log\left(\frac{1-r^2_{\tau}}{1-r^2_{0}}\right) - \frac{r_{\tau}}{2}\log\left(\frac{1+r_{\tau}}{1-r_{\tau}}\right) + \frac{r_{0}}{2}\log\left(\frac{1+r_{0}}{1-r_{0}}\right)+\beta (z_{0}-z_{\tau}),
\end{equation}
where $r_{\tau}$ has been defined as in Example II, the length of the Bloch vector. If we consider the special case where $x_{0}=y_{0}=z_{0}=0$ (implying in turn $r_{\tau}=|z_{\tau}|$), Eq.~\eqref{diff} is simplified as follows:
\begin{equation}
\bm{\Sigma}_{\tau}
=-\frac{1+|z_{\tau}|}{2}\log\left([1+|z_{\tau}|]\mrm{e}^{\beta z_{\tau}}\right)- \frac{1-|z_{\tau}|}{2}\,\log\left([1-|z_{\tau}|]\mrm{e}^{\beta z_{\tau}}\right)\ .
\end{equation}
Figure \ref{fig:1} depicts this quantity for $\beta=0.1$ and $\epsilon=\lambda=1$, which explicitly shows that $\bm{\Sigma}_{\tau}\leqslant0$ in a certain time interval.

On the basis of these two examples, one may conclude that either the second law of thermodynamics can be violated by physically legitimate dynamical maps, or 
a more careful formulation of the second law should be given. 
The latter possibility seems more natural in particular because a general 
statement has been proven in Refs. \cite{book:Peres,Esposito,Sagawa} considering explicitly both system $S$ and bath $B$ in the entropy balance.  
Specifically, it has been shown that
\begin{equation}
\label{true2law}
\Delta \mathds{S}_{S,\tau}+ \Delta \mathds{S}_{B,\tau}\geqslant 0,
\end{equation}
where $\Delta \mathds{S}_{S,\tau}:= \mathds{S}_{S,\tau}-\mathds{S}_{S,0}$ and $\Delta \mathds{S}_{B,\tau}:= \mathds{S}_{B,\tau}-\mathds{S}_{B,0} $.
This inequality is valid provided that the initial state of the composite system $SB$ is factorized, without further particular restrictions on the reduced dynamics of $S$  
or $B$. Indeed, under this condition one can see that
\begin{equation}
\Delta \mathds{S}_{S,\tau}+ \Delta \mathds{S}_{B,\tau}= \mathds{S}(\varrho_{SB,\tau}|\hskip-0.5mm|\varrho_{S,\tau}\otimes \varrho_{B,\tau}) \geqslant 0.
\end{equation} 
In this respect,  Eq.~\eqref{true2law} should be considered as 
a general formulation of the second law. 

Conversely, we have shown that the validity of $\bm{\sigma}_{\tau} \geqslant 0$ is subject to further dynamical constraints. Heuristically, one can think of obtaining $\bm{\sigma}_{\tau} \geqslant 0$ as a particular case of relation \eqref{true2law} in three steps. First, one should assume that relation \eqref{true2law} holds 
also in a differential form, $\partial_{\tau} \mathds{S}_{S,\tau}+ \partial_{\tau} \mathds{S}_{B,\tau}\geqslant 0$. Moreover, since the bath is usually considered in thermal equilibrium at inverse temperature $\beta$, 
one can use the relation $\partial_{\tau} \mathds{S}_{B,\tau}= \beta \partial_{\tau} \mathds{Q}_{B,\tau}$. Finally, the heat flux of the bath is basically related to the heat flux of the system as $\partial_{\tau} \mathds{Q}_{B,\tau}=-\partial_{\tau} \mathds{Q}_{S,\tau}$. These assumptions, although reasonable, can be violated if the system and the bath are strongly coupled and correlated. Thus one should not consider $\bm{\sigma}_{\tau}\geqslant0$ as an \textit{a priori} valid formulation of the second law.

\section*{Example III: Dephasing qubit}

\label{sec: example3}

In this section we use the approach presented in Ref. 
\cite{Alipour} (and mentioned after Eq.~\ref{ineq}), which explicitly considers the presence of a second system interacting with the one of thermodynamic interest---without eliminating it by any effective procedure as in the usual weak-coupling limit. In this framework, $\Delta \mathds{S}_{B,\tau}$ is explicitly computed, together with $\Delta \mathds{S}_{S,\tau}$, and the previous inequality~\eqref{true2law} naturally arises. Instead, the three assumptions mentioned after Eq.~\eqref{true2law}
that relate $\bm{\sigma}_{\tau}$ and Eq.~\eqref{true2law} are in general violated, as shown in the following example. 

Consider a total Hamiltonian given by $H_{\mathrm{tot}}=H_S+H_B+H_{\mathrm{int}}$ with
\begin{equation}
H_S=\frac{\omega_{0}}{2}\sigma_z, \quad H_B=\sum_{k=1}^{\infty}\omega_k \mathrm{a}^\dag_k \mathrm{a}_k\ ,\quad H_{\mathrm{int}}=\lambda\sigma_z\otimes\sum_{k=1}^{\infty}\big(f_k^*\mathrm{a}_k+f_k\mathrm{a}^{\dag}_k\big),
\end{equation}
where $\mathrm{a}_k$ is the bosonic annihilation operator of mode $k$, satisfying the canonical commutation relations $[\mathrm{a}_k,\mathrm{a}^\dag_l]=\delta_{kl}$, and the complex parameters $f_k$ are such that $\sum_{k=1}^{\infty}|f_k|^2<\infty$. We assume that the initial state of the total system can be written as $\varrho_{SB,0}=\varrho_{S,0}\otimes \varrho_{B}^{(\beta)}$, where $\varrho_{S,0}$ is the initial state of the qubit and $\varrho_{B}^{(\beta)}$ is the Gibbs state of the thermal bath at inverse temperature $\beta$,
\begin{equation}
\label{Sst}
\varrho_{S,0}=\sum_{\ell,\ell'=0}^1\varrho_{\ell\ell'}\vert\ell\rangle\langle\ell'\vert,\quad \varrho_B^{(\beta)}=\frac{\mathrm{e}^{-\beta\sum_k\omega_k\mathrm{a}^\dag_k\rm{a}_k}}{\mathrm{Tr}\left[\mathrm{e}^{-\beta\sum_k\omega_k \mathrm{a}^\dag_k\mathrm{a}_k}\right]}\ ,\qquad
\sigma_z\ket{\ell}= (-)^\ell\ket{\ell}.
\end{equation}  
The dynamics of the total system can be analytically solved (see Ref. \cite{Alipour} for details) and by partial tracing one can obtain the reduced density matrices of the two subsystems at any time, $\varrho_{S,\tau}$ and $\varrho_{B,\tau}$. One can quantify the correlations between $S$ and $B$ through the operator $\chi_{\tau}:=\varrho_{SB,\tau} -\varrho_{S,\tau}\otimes\varrho_{B,\tau}$, which plays a prominent role in 
the approach of Ref. \cite{Alipour}. 
This can be seen from the variation of the total energy $\mathds{U}_{\mrm{tot}}:=\mathrm{Tr}\left[H_{\mrm{tot}}\varrho_{SB,\tau}\right]$ as
\begin{eqnarray}
\nonumber
\partial_{\tau}\mathds{U}_{\mrm{tot}}&=&\mathrm{Tr}\left[H_{\mrm{tot}}\,\partial_{\tau}\varrho_{S,\tau}\otimes \varrho_{B\tau} \right]+ \mathrm{Tr}\left[H_{\mrm{tot}}\,\varrho_{S,\tau}\otimes\partial_{\tau} \varrho_{B,\tau} \right]\,
+\, \mathrm{Tr}\left[H_{\mrm{tot}}\,\partial_{\tau} \chi_{\tau}\right]\\ 
&=& \mrm{Tr}\left[ \partial_{\tau}\varrho_{S,\tau} \,H'_{S,\tau} \right]\,+\, \mrm{Tr}\left[ \partial_{\tau}\varrho_{B,\tau} \,H'_{B,\tau} \right]+ \mathrm{Tr}[H_{\mathrm{int}}\,\partial_{\tau}\chi_{\tau}]=0\ ,
\label{der}
\end{eqnarray}
where a modified Hamiltonian $H'_{a,\tau}$ ($a\in\{S,B\}$) has been defined for each subsystem as 
\begin{equation}
H'_{S,\tau}:=H_{S}+\mathrm{Tr}_{B}\left[\varrho_{B,\tau}H_{\mathrm{int}}\right],
\end{equation}
(and similarly for $H'_{B,\tau}$) and the last equality in Eq.~\eqref{der} holds because the global unitary evolution is generated by $H_{\mathrm{tot}}$. 
These are the same Hamiltonians that one finds in the evolution equation for the reduced density matrices
\begin{align}
\partial_{\tau}\varrho_{S,\tau}=&-i\,[H'_{S,\tau},\varrho_{S,\tau}] -i\,\mathrm{Tr}_B\left[H_{\mathrm{int}},\chi_\tau\right] ,\label{PT3}
\\
\partial_{\tau}\varrho_{B,\tau}=&-i\,[H'_{B,\tau},\varrho_{B,\tau}] -i\,\mathrm{Tr}_S\left[H_{\mathrm{int}},\chi_\tau\right]\ ,\label{PT4}
\end{align}
where the second terms (after the commutators) account for dynamical correlations between the two systems and which, in the weak-coupling limit, would give rise to a dissipative contribution to the master equation for the reduced dynamics.
One can then interpret the first two terms in the second line of Eq.~\eqref{der} as the heat exchanged by $S$, respectively $B$, because they are similar to Eq.~\eqref{heat} with the effective Hamiltonians substituting $H_S$ and $H_B$, 
\begin{equation}
\partial_{\tau}\mathds{Q}_{S,\tau}:= \mathrm{Tr}[\partial_{\tau}\varrho_{S,\tau} \,H'_{S,\tau} ], \quad \partial_{\tau}\mathds{Q}_{B,\tau}:= \mathrm{Tr}[\partial_{\tau}\varrho_{B,\tau} \,H'_{B,\tau} ].
\end{equation} 
Accordingly, the last term in Eq.~(\ref{der}) can thus be associated with the variation of the energy stored in the correlations, which is called the 
binding energy $\mathds{U}_{\chi,\tau}:=\mathrm{Tr}[\chi_{\tau}\,H_{\mathrm{int}}]$.

Therefore, one of the three steps in the comments after Eq.~\eqref{true2law}
is invalid in the presence of correlations between the subsystems because the heat balance now reads 
\begin{equation}
\label{heatbalance}
\partial_{\tau}\mathds{Q}_{S,\tau}+\partial_{\tau}\mathds{Q}_{B,\tau}=-\partial_{\tau}\mathds{U}_{\chi,\tau}.
\end{equation}
Moreover, a second assumption is also found to be unwarranted. Indeed, even though Eq.~\eqref{true2law} holds because the relative entropy is positive,  
the differential statement
\begin{equation}
\partial_{\tau}\mathds{S}_{S,\tau} + \partial_{\tau}\mathds{S}_{B,\tau} \geqslant 0
\end{equation}
does not remain valid in general. 

In the following, we present the explicit expressions of the entropy variations and heat fluxes for both system and bath, computed in Ref. \cite{Alipour} using the previous approach. This highlights how the assumptions can be violated in a physically meaningful model.

Concerning qubit $S$, one finds that $\partial_{\tau}\mathds{Q}_{S,\tau}=0$, 
whereas the entropy is
\begin{equation}
\partial_{\tau}\mathds{S}_{S,\tau}=-\frac{1}{2}\log\!\left(\frac{1+r_{S,\tau}}{1-r_{S,\tau}}\right)\,\partial_{\tau} r_{S,\tau}=\lambda^{2}\frac{16|\varrho_{01}|^2 \,\mathrm{e}^{-16\lambda^2\Gamma_{\tau}}}{r_{S,\tau}}\log\!\left(\frac{1+r_{S,\tau}}{1-r_{S,\tau}}\right)\partial_{\tau}\Gamma_{\tau},
\end{equation}
where
\begin{equation}
\Gamma_{\tau}=\int_0^{\infty}\mathrm{d}\omega \frac{|f(\omega)|^2}{\omega^2} \coth(\beta\omega/2)\sin^2(\omega \tau/2)\ ,\quad 
r_{S,\tau}=\sqrt{1-4\left(\varrho_{00}\varrho_{11}-\mathrm{e}^{-16\lambda^2\Gamma_{\tau}}|\varrho_{01}|^2\right)}.
\end{equation}
In writing $\Gamma_{\tau}$, the continuum limit has been taken and the sum over the bath modes $\sum_k |f_k|^2$ has been recast into the integral $\int_0^{\infty}|f(\omega)|^2\,\mathrm{d}\omega$. Concerning the bath quantities, one has
\begin{equation}
\partial_{\tau}\mathds{Q}_{B,\tau}=4\lambda^2\big(1-\average{\sigma_z}^2\big)\,\partial_{\tau}\Delta_{\tau}\ ,\quad
\partial_{\tau}\mathds{S}_{B,\tau}=4 \beta\lambda^2\big(1-\average{\sigma_z}^2\big)\,\partial_{\tau}\Delta_{\tau} + 
O(\lambda^3),
\end{equation}
where
\begin{equation}
\Delta_{\tau}=\int_0^{\infty} \frac{|f(\omega)|^2}{\omega} \sin^2(\omega \tau/2)\,\mathrm{d}\omega\ .
\end{equation}

It is evident that $\partial_{\tau}\mathds{S}_{B,\tau}= \beta\partial_{\tau}\mathds{Q}_{B,\tau}$, up to leading order in the coupling constant, so that the hypothesis of a thermal bath almost in equilibrium seems to be robust. However, one obtains that $\partial_{\tau}\mathds{Q}_{S,\tau}\neq \partial_{\tau}\mathds{Q}_{B,\tau}$ because the first one is identically vanishing whereas the latter is not. As already mentioned, this is possible due to the correlations between the subsystems that can store and exchange energy, effectively acting as a third subsystem \cite{Alipour}. The third hypothesis can be also violated. Indeed, one can show that $\partial_{\tau}\mathds{S}_{S,\tau} + \partial_{\tau}\mathds{S}_{B,\tau}$ possibly becomes negative even though its integral is always positive. The sign of $\partial_{\tau}\mathds{S}_{B,\tau}$ is equal to the sign of $\partial_{\tau} \Delta_{\tau}$; whereas the sign of $\partial_{\tau}\mathds{S}_{S,\tau}$ depends on $\partial_{\tau} \Gamma_{\tau}$. One can see that $\partial_{\tau} \Gamma_{\tau}<0$, which corresponds to an essentially non-Markovian dynamics for $S$, by choosing a super-Ohmic spectral density
\begin{equation}
|f(\omega)|^2= \frac{\omega^{s}}{\omega_{\mathrm{c}}^{s-1}}\,\mrm{e}^{-\omega/\omega_{\mathrm{c}}},
\end{equation}
with $s>s_{\mathrm{cr}}(\beta)$ ($\omega_{\mathrm{c}}$ is a cutoff frequency). The critical ohmicity parameter $s_{\mathrm{cr}}$ at zero temperature is $2$, but it becomes $3$ in the infinite temperature limit \cite{dephasing}. Indeed, for high temperature one can expand the hyperbolic cotangent in $\Gamma_{\tau}$ and
\begin{equation}
\partial_{\tau} \Gamma_{\tau}\simeq \frac{1}{2\beta}\widetilde{\Gamma}(s-1)\left[1+(\omega_{\mathrm{c}} \tau)^2\right]^{-\frac{s-1}{2}}\sin[(s-1)\arctan(\omega_{\mathrm{c}} \tau)],
\end{equation}
where $\widetilde{\Gamma}$ is the Euler gamma function.
Moreover, one can see that $\partial_{\tau}\Delta_{\tau}<0$ if $s>1$, because
\begin{equation}
\partial_{\tau} \Delta_{\tau}=  \frac{\omega_{\mathrm{c}}^2}{2}\,\widetilde{\Gamma}(s+1)\left[1+(\omega_{\mathrm{c}} \tau)^2\right]^{-\frac{s+1}{2}}\sin[(s+1)\arctan(\omega_{\mathrm{c}} \tau)].
\end{equation}
Thus for $s=4$ and at sufficiently high temperature one can find $\partial_{\tau} \mathds{S}_{S,\tau}<0$ and $\partial_{\tau} \mathds{S}_{B,\tau}<0$ simultaneously. 
This happens when $\pi/3<\arctan(\omega_{\mathrm{c}} \tau)<\pi/2$. 

This example explicitly shows that, in general, the statement \eqref{true2law} of the second law is not equivalent to $\bm{\sigma}_{\tau}\geqslant 0$. Hence a violation of the latter inequality should not be interpreted as unphysical. 
Note that, while the expressions for heat and work are those proposed in~\cite{Alicki-79, Spohn}, the Hamiltonian operators appearing in them are effectively redefined to take into account the interaction between system and environment. In this sense our perspective is similar in spirit to the ones in~\cite{Mahler, Esposito2}, the concrete operative definitions of heat and work being different, though.

\section*{Conclusions}
\label{sec:conc}

In this paper, we have studied the entropy production, or differential entropy rate, in open quantum system undergoing non-Markovian time-evolutions that extend those obtained via the weak-coupling limit and the Markovian approximation. In this framework, the open system dynamics is dealt with by eliminating the bath degrees of freedom; hence, bath entropy variations cannot contribute to the entropy balance.
In particular, we have shown that the class of so-called essentially non-Markovian dynamics is compatible with a negative entropy production. Moreover, also the integrated entropy production for the open system alone can be negative if the asymptotic state is thermal, but not invariant at finite times, a fact impossible when the dynamics satisfies the semigroup composition law.

Unlike when it is the lack of complete-positivity that leads to a negative entropy production, we have explicitly shown that such a phenomenon can also be due to completely-positive (thus physically legitimate), but non-Markovian dynamics. This outcome cannot be interpreted as a violation of the second law of thermodynamics. On the contrary, it suggests a more standard approach: in the presence of a non-Markovian reduced dynamics, a proper formulation of the second law of thermodynamics requires the bath to be explicitly considered instead of being effectively eliminated by weak-coupling limit techniques. Including the bath entropy variations to those of the entropy of the open subsystem, one then obtains the positivity of the entropy balance.


\section*{Acknowledgements}

S.M. thanks S. Bacchi, L. Curcuraci, G. Gasbarri, and J. Goold for useful comments on the manuscript.
F.B. and S.M. gratefully acknowledge financial support from the University of Trieste research fund FRA-2015.

\section*{Author contributions statement}

S.M. conceived the main theoretical ideas, S.A., F.B., R.F., and A.T.R. discussed and analyzed the results.  All authors contributed in writing and revising the manuscript. 

\section*{Additional information}

There are no \textbf{competing financial interests}. 

\newpage

\section*{Supplementary Information: Sign of the entropy production in Example I}

From Eq.~(18) in the main text it seems that the sign of $\bm{\sigma}_{\tau}$ depends both on $\gamma_{\tau}$ and on the sign of the expression within the square brackets. In this section we prove that the latter is always positive. Let us rewrite the entropy production in a more convenient way as
$$
\bm{\sigma}_{\tau} = \gamma_{\tau}\coth(\beta\omega/2)\,\mrm{e}^{-2\Gamma_{\tau}} [A+B+C],
$$
where
$$
A= \frac{x^2_{0}+y^2_{0}}{4r_{\tau}}\log\left(\frac{1+r_{\tau}}{1-r_{\tau}}\right), \quad
B= \frac{z_{0}+|z_{\infty}|}{2}\log\left(\frac{1+r_{\infty}}{1-r_{\infty}}\right), \quad
C= \frac{(z_{0}+|z_{\infty}|)z_{\tau}}{2r_{\tau}}\log\left(\frac{1+r_{\tau}}{1-r_{\tau}}\right).
$$
Note that $A$ is always nonnegative whereas $B$ and $C$ can be either positive or negative. We show in the following that $A+B+C$ is nevertheless positive, distinguishing different situations. 
\begin{enumerate}
\item 
 If $z_{0}+|z_{\infty}|\leqslant 0$, then $z_{\tau}\leqslant -|z_{\infty}|$, because $z_{\tau}+|z_{\infty}|= \mrm{e}^{-2\Gamma_{\tau}}(z_{0}+|z_{\infty}|)$ and 
$$
|r_{\tau}|\geqslant |z_{\tau}|\geqslant |z_{\infty}|= r_{\infty}.
$$
Hence $B+C\geqslant 0$, because 
\begin{equation*}
r_{\tau}\log\left(\frac{1+r_{\infty}}{1-r_{\infty}}\right)- |z_{\tau}|\log\left(\frac{1+r_{\tau}}{1-r_{\tau}}\right) \leqslant r_{\tau}\log\left(\frac{1+r_{\infty}}{1-r_{\infty}}\right)- r_{\infty}\log\left(\frac{1+r_{\tau}}{1-r_{\tau}}\right)\leqslant 0,
\end{equation*}
where the last inequality holds because the function
\begin{equation*}
f(x)=\frac{1}{x}\log\left(\frac{1+x}{1-x}\right), 
\end{equation*}
is monotonically increasing for $0<x<1$. This can be seen from the first derivative,
\begin{equation*}
f^{\prime}(x)= \frac{1}{x^2}\left[\frac{2x}{1-x^2}-\log\left(\frac{1+x}{1-x}\right)\right],
\end{equation*}
which is is always positive because
\begin{equation*}
\frac{2x}{1-x^2}- |\log(1+x)|-|\log(1-x)| \geqslant \frac{2x}{1-x^2}- \frac{x}{2}\frac{2+x}{1+x}- \frac{x}{2}\frac{2-x}{1-x}=\frac{x^3}{1-x^2}\geqslant0,
\end{equation*}
in which the following inequalities have been used \cite{log-ineq}:
\begin{equation*}
\frac{2x}{2+x}\leqslant|\log(1+x)|\leqslant \frac{x}{2}\frac{2+x}{1+x}, \qquad \frac{2x}{2-x}\leqslant|\log(1-x)|\leqslant \frac{x}{2}\frac{2-x}{1-x}.
\end{equation*}

\item In the case $z_{0}+|z_{\infty}|\geqslant 0$, we need to distinguish different situations. 
  \begin{itemize}
   \item  First, 
if $z_{\tau}\geqslant0$, then $B$ and $C$ are both positive. 
   \item  If instead $-|z_{\infty}|\leqslant z_{\tau}\leqslant0$ and $r_{\tau}\leqslant r_{\infty}$, then $B+C$ is positive because
\begin{equation*}
\log\left(\frac{1+r_{\infty}}{1-r_{\infty}}\right)- \frac{|z_{\tau}|}{r_{\tau}}\log\left(\frac{1+r_{\tau}}{1-r_{\tau}}\right) \geqslant \log\left(\frac{1+r_{\infty}}{1-r_{\infty}}\right)- \log\left(\frac{1+r_{\tau}}{1-r_{\tau}}\right)\geqslant 0.
\end{equation*}
   \item  The last possibility is $-|z_{\infty}|\leqslant z_{\tau}\leqslant0$ and $r_{\tau}\geqslant r_{\infty}$. In this case, $B$ is positive and the following inequality holds: 
\begin{equation*}
x^2_{0}+y^2_{0}\geqslant(|z_{\infty}|-z_{\tau})(z_{0}+|z_{\infty}|)\geqslant 0.
\end{equation*}
As a consequence, $A+C\geqslant 0$,
\begin{equation*}
x^2_{0}+y^2_{0}-2|z_{\tau}|(z_{0}+|z_{\infty}|)\geqslant x^2_{0}+y^2_{0}-(|z_{\infty}|+|z_{\tau}|)(z_{0}+|z_{\infty}|)\geqslant0.
\end{equation*}
\end{itemize}
\end{enumerate}
Summarizing, the expression in the square brackets $[A+B+C]$ is always nonnegative and the sign of the entropy production is only determined by $\gamma_{\tau}$.


\begin{thebibliography}{1}
\expandafter\ifx\csname url\endcsname\relax
  \def\url#1{\texttt{#1}}\fi
\expandafter\ifx\csname urlprefix\endcsname\relax\def\urlprefix{URL }\fi
\expandafter\ifx\csname doiprefix\endcsname\relax\def\doiprefix{DOI }\fi
\providecommand{\bibinfo}[2]{#2}
\providecommand{\eprint}[2][]{\url{#2}}

\bibitem{Alicki_book} 
\bibinfo{author}{Alicki, R.} \& \bibinfo{author}{Lendi, K.} 
\newblock\bibinfo{book}{\bibinfo{title}{\emph{Quantum Dynamical Semigroups and Applications}.}} 
\newblock\bibinfo{publisher}{Springer, Berlin, 2007}.

\bibitem{Book:deGroot} 
\bibinfo{author}{de Groot, S.R.} \& \bibinfo{author}{Mazur, P.} 
\newblock\bibinfo{book}{\bibinfo{title}{\emph{Non-Equilibrium Thermodynamics.}}} 
\newblock\bibinfo{pubisher}{Dover
Publications, New York, 1984}.

\bibitem{Alicki-79}
\bibinfo{author}{Alicki, R.} 
\newblock\bibinfo{title}{The quantum open system as a model of the heat engine.}
\newblock{\emph{J. Phys. A: Math. Gen.}} 
\textbf{\bibinfo{volume}{12}}, \bibinfo{pages}{103} (\bibinfo{year}{1979}).
%

\bibitem{Spohn} 
\bibinfo{author}{Spohn H.} \&  \bibinfo{author}{Lebowitz, J.L.}
\newblock\bibinfo{title}{Irreversible thermodynamics for quantum systems weakly coupled to thermal reservoirs.}
\newblock{\emph{Adv. Chem. Phys.}} 
\textbf{\bibinfo{volume}{38}}, \bibinfo{pages}{109} (\bibinfo{year}{1979}).
%

\bibitem{entroSpohn} 
\bibinfo{author}{Spohn, H.}
\newblock\bibinfo{title}{Entropy production for quantum dynamical semigroups}
\newblock{\emph{J. Math. Phys.}} 
\textbf{\bibinfo{volume}{19}}, \bibinfo{pages}{1227} (\bibinfo{year}{1978}).
%

\bibitem{review} 
\bibinfo{author}{Breuer, H.-P.}, \bibinfo{author}{Laine, E.M.}, \bibinfo{author}{Piilo, J.} \& \bibinfo{author}{Vacchini, B.}
\newblock\bibinfo{title}{Colloquium: Non-Markovian dynamics in open quantum systems.}
\newblock{\emph{Rev. Mod. Phys.}}
\textbf{\bibinfo{volume}{88}}, \bibinfo{pages}{021002} (\bibinfo{year}{2016}).
%

\bibitem{RHPreview}
\bibinfo{author}{Rivas, A.}, \bibinfo{author}{Huelga, S.F.} \& \bibinfo{author}{Plenio, M.B.}
\newblock\bibinfo{title}{Quantum non-Markovianity: characterization, quantification and detection.}
\newblock{\emph{Rep. Prog. Phys.}} \textbf{\bibinfo{volume}{77}}, \bibinfo{pages}{094001} (\bibinfo{year}{2014})
%

\bibitem{deVega}
\bibinfo{author}{de Vega, I} \& \bibinfo{author}{Alonso, D.} 
\newblock\bibinfo{title}{Dynamics of non-Markovian open quantum systems.}
\newblock{\emph{Rev. Mod. Phys.}}
\textbf{\bibinfo{volume}{89}}, 
\bibinfo{pages}{015001} (\bibinfo{year}{2017}) 
%

\bibitem{RHP} 
\bibinfo{author}{Rivas, A.}, \bibinfo{author}{Huelga, S.F.} \& \bibinfo{author}{Plenio, M.B.}
\newblock\bibinfo{title}{Entanglement and non-Markovianity of quantum evolutions.}
\newblock{\emph{Phys. Rev. Lett.}}
\textbf{\bibinfo{volume}{105}}, \bibinfo{pages}{050403} (\bibinfo{year}{2010}).
%

\bibitem{Chruscinski}  
\bibinfo{author}{Chru\'{s}ci\'{n}ski, D.} \& \bibinfo{author}{Kossakowski, A.} 
\newblock\bibinfo{title}{Non-Markovian quantum dynamics: Local versus non-local.}
\newblock{\emph{Phys. Rev. Lett.}} 
\textbf{\bibinfo{volume}{104}}, \bibinfo{pages}{070406} (\bibinfo{year}{2010}).
%

\bibitem{Fazio} 
\bibinfo{author}{Mukherjee, V.}, \bibinfo{autor}{Giovannetti, V.}, \bibinfo{author}{Fazio, R.}, \bibinfo{auhtor}{Huelga, S.F.}, \bibinfo{author}{Calarco, T.} \& \bibinfo{Montangero, S.} 
\newblock\bibinfo{title}{Efficiency of quantum controlled non-Markovian thermalization.}
\newblock{\emph{New J. Phys.}} 
\textbf{\bibinfo{volume}{17}}, \bibinfo{pages}{063031} (\bibinfo{year}{2015}).
%

\bibitem{Bylicka} 
\bibinfo{author}{Bylicka, B.}, \bibinfo{author}{Tukiainen, M.}, \bibinfo{author}{Chru\'{s}ci\'{n}ski, D.}, \bibinfo{author}{Piilo, J.} \& \bibinfo{author}{Maniscalco, S.} 
\newblock\bibinfo{title}{Thermodynamic power of non-Markovianity.}
\newblock{\emph{Sci. Rep.}} 
\textbf{\bibinfo{volume}{6}}, \bibinfo{pages}{27989} (\bibinfo{year}{2016}).


\bibitem{Argentieri1} 
\bibinfo{author}{Argentieri, G.}, \bibinfo{author}{Benatti, F.}, \bibinfo{author}{Floreanini, R.} \& \bibinfo{author}{Pezzutto, M.} 
\newblock\bibinfo{title}{Violations of the second law of thermodynamics by a non-completely positive dynamics.}
\newblock{\emph{Eur. Phys. Lett.}}
\textbf{\bibinfo{volume}{107}}, \bibinfo{pages}{50007} (\bibinfo{year}{2014}).
%

\bibitem{Argentieri2} 
\bibinfo{author}{Argentieri, G.}, \bibinfo{author}{Benatti, F.}, \bibinfo{author}{Floreanini, R.} \& \bibinfo{author}{Pezzutto, M.}  
\newblock\bibinfo{title}{Complete positivity and thermodynamics in a driven open quantum system.}
\newblock{\emph{J. Stat. Phys.}} 
\textbf{\bibinfo{volume}{159}}, \bibinfo{pages}{1127} (\bibinfo{year}{2015}).
%


\bibitem{book:Peres} 
\bibinfo{author}{Peres, A.} 
\newblock\bibinfo{book}{\bibinfo{title}{\emph{Quantum Theory: Concepts and Methods.}}} \newblock\bibinfo{publisher}{Kluwer Academic Publishers, New York, 1995}.
%

\bibitem{Esposito} 
\bibinfo{author}{Esposito, M.}, \bibinfo{author}{Lindenberg, K.} \& \bibinfo{author}{Van den Broeck, C.}  
\newblock\bibinfo{title}{Entropy production as correlation between system and reservoir.
}
\newblock{\emph{New J. Phys.}}
\textbf{\bibinfo{volume}{12}}, \bibinfo{pages}{013013} (\bibinfo{year}{2010}).
%


\bibitem{Sagawa} 
\bibinfo{author}{Sagawa, T.} 
\newblock\bibinfo{title}{Second law-like inequalities with quantum relative entropy: An introduction.}
\newblock\bibinfo{book}{\bibinfo{title}{In \emph{Lectures on Quantum Computing, Thermodynamics and Statistical Physics.} Edited by M. Nakahara and S. Tanaka.}} 
\newblock\bibinfo{pubisher}{World Scientific, Singapore, 2012;}
\newblock{\emph{arXiv:1202.0983v3 [cond-mat.stat-mech]}} 
(\bibinfo{year}{2014}).
%


\bibitem{Modi} 
\bibinfo{author}{Pollok, F.A.}, \bibinfo{author}{Rodriguez-Rosario, C.}, \bibinfo{author}{Frauenheim, T.}, \bibinfo{author}{Paternostro, M.} \& \bibinfo{author}{Modi, K.} 
\newblock\bibinfo{title}{Complete framework for efficient characterisation of non-Markovian processes.}
\newblock{\emph{arXiv:1512.00589v2 [quant-ph]}} 
(\bibinfo{year}{2016}).
%

\bibitem{asymptotic} 
\bibinfo{author}{Chr\'{u}sci\'{n}ski, D.}, \bibinfo{author}{Kossakowski, A.} \& 
\bibinfo{author}{Pascazio, S.} 
\newblock\bibinfo{title}{Long-time memory in non-Markovian evolutions.}
\newblock{\emph{Phys. Rev. A}} 
\textbf{\bibinfo{volume}{81}}, \bibinfo{pages}{032101} (\bibinfo{year}{2010}).
%


\bibitem{book:Nielsen} 
\bibinfo{author}{Nielsen, M.A.} \& \bibinfo{author}{Chuang, I.L.} 
\newblock\bibinfo{book}{\bibinfo{title}{\emph{Quantum Computation and Quantum Information.}}} \newblock\bibinfo{publisher}{Cambridge University Press, Cambridge, 2000}.

\bibitem{Alipour} 
\bibinfo{author}{Alipour, S.}, \bibinfo{author}{Benatti, F.}, \bibinfo{author}{Bakhshinezhad, F.}, \bibinfo{author}{Afsary, M.}, \bibinfo{author}{Marcantoni, S.} \&  
\bibinfo{author}{Rezakhani, A.T.} 
\newblock\bibinfo{title}{Correlations in quantum thermodynamics: Heat, work, and entropy production.}
\newblock{\emph{Sci. Rep.}}
\textbf{\bibinfo{volume}{6}}, \bibinfo{pages}{35568} (\bibinfo{year}{2016}).
%

\bibitem{degree} 
\bibinfo{author}{Chru\'{s}ci\'{n}ski, D.} \& \bibinfo{author}{Maniscalco, S.}
\newblock\bibinfo{title}{On the degree of non-Markovianity of quantum evolution.}
\newblock{\emph{Phys. Rev. Lett.}} 
\textbf{\bibinfo{volume}{112}}, \bibinfo{pages}{120404} (\bibinfo{year}{2014}).
%


\bibitem{positive}  
\bibinfo{author}{M\"uller-Hermes, A.} \& \bibinfo{author}{Reeb, D.}
\newblock\bibinfo{title}{Monotonicity of the quantum relative entropy under positive maps.}
\newblock{\emph{Ann. Henri Poincar\'{e}}} 
\textbf{\bibinfo{volume}{18}}, \bibinfo{pages}{1777}
(\bibinfo{year}{2017}). 
%

\bibitem{commutative} 
\bibinfo{author}{Chru\'{s}ci\'{n}ski, D.} \& \bibinfo{author}{Kossakowski, A.}
\newblock\bibinfo{title}{Markovianity criteria for quantum evolution.}
\newblock{\emph{J. Phys. B: At. Mol. Opt. Phys.}}
\textbf{\bibinfo{volume}{45}}, \bibinfo{pages}{154002} (\bibinfo{year}{2012}).
%


\bibitem{dephasing} 
\bibinfo{author}{Haikka, P.}, \bibinfo{author}{Johnson, T.H.} \& \bibinfo{author}{Maniscalco, S.}
\newblock\bibinfo{title}{Non-Markovianity of local dephasing channels and time-invariant discord.}
\newblock{\emph{Phys. Rev. A}}
\textbf{\bibinfo{volume}{87}}, \bibinfo{pages}{010103(R)} (\bibinfo{year}{2013}).
%

\bibitem{Mahler}
\bibinfo{author}{Teifel, J.} \& \bibinfo{author}{Mahler, G.}  
\newblock\bibinfo{title}{Autonomous modular quantum systems: Contextual Jarzynski relations.}
\newblock{\emph{Phys. Rev. E}}
\textbf{\bibinfo{volume}{83}}, \bibinfo{pages}{041131} (\bibinfo{year}{2011}).
%

\bibitem{Esposito2}
\bibinfo{author}{Esposito, M.}, \bibinfo{author}{Ochoa, M.A.} \& \bibinfo{author}{Galperin, M.}  
\newblock\bibinfo{title}{Nature of heat in strongly coupled open quantum systems.}
\newblock{\emph{Phys. Rev. B}}
\textbf{\bibinfo{volume}{92}}, \bibinfo{pages}{235440} (\bibinfo{year}{2015}).
%


\end{thebibliography}

\begin{thebibliography}{2}
\expandafter\ifx\csname url\endcsname\relax
  \def\url#1{\texttt{#1}}\fi
\expandafter\ifx\csname urlprefix\endcsname\relax\def\urlprefix{URL }\fi
\expandafter\ifx\csname doiprefix\endcsname\relax\def\doiprefix{DOI }\fi
\providecommand{\bibinfo}[2]{#2}
\providecommand{\eprint}[2][]{\url{#2}}

\bibitem{log-ineq} 
\bibinfo{author}{Tops{\o}e, F.} 
\newblock\bibinfo{title}{Some Bounds for the Logarithmic Function.} 
\newblock{in: Cho, Y.J., Kim, J.K. \& Dragomir, S.S.}, 
\newblock\bibinfo{book}{\bibinfo{title}{\emph{Inequality Theory and Applications.}}}
\newblock\bibinfo{publisher}{Nova Science Publishers, New York, 2007}.

\end{thebibliography}
\end{document}